\documentclass{aa}  
\usepackage{graphicx}
\usepackage{subfig}
\usepackage{txfonts}
\usepackage{natbib}
\usepackage{epstopdf}
\usepackage[breaklinks=true]{hyperref} 
\hypersetup{colorlinks=true,urlcolor=blue,citecolor=blue,pdfborder= 0 0 0}
\bibpunct{(}{)}{;}{a}{}{,}
\usepackage{color}

\begin{document} 

\title{Spiral arms and disc stability in the Andromeda galaxy}

   \author{P. Tenjes\inst{1,2},
          T. Tuvikene\inst{1},
          A. Tamm\inst{1},
          R. Kipper\inst{1},
          \and 
          E. Tempel\inst{1,3}
          }
   \institute{
          Tartu Observatory, Observatooriumi 1, 61602 T\~oravere, Estonia\\
                  \email{peeter.tenjes@to.ee}
          \and
          Institute of Physics, University of Tartu, W. Ostwaldi 1, 50411 Tartu, Estonia
          \and
                  Leibniz-Institut f\"ur Astrophysik Potsdam (AIP), An der Sternwarte 16, 14482 Potsdam, Germany}

   \date{Received November 2, 2016; accepted January 18, 2017}

   \authorrunning{P. Tenjes et al.}
   \titlerunning{Spirals of M\,31 and disc stability}
 
   \abstract
    {}
   {Density waves are often considered as the triggering mechanism of star formation in spiral galaxies. Our aim is to study relations between different star formation tracers (stellar UV and near-IR radiation and emission from \ion{H}{I}, CO, and cold dust) in the spiral arms of M\,31, to calculate stability conditions in the galaxy disc, and to draw conclusions about possible star formation triggering mechanisms.}
   {We selected fourteen spiral arm segments from the de-projected data maps and compared emission distributions along the cross sections of the segments in different datasets to each other, in order to detect  spatial offsets between young stellar populations and the star-forming medium. By using the disc stability condition as a function of perturbation wavelength and distance from the galaxy centre, we calculated the effective disc stability parameters and the least stable wavelengths at different distances. For this we used a mass distribution model of M\,31 with four disc components (old and young stellar discs, cold and warm gaseous discs) embedded within the external potential of the bulge, the stellar halo, and the dark matter halo. Each component is considered to have a realistic finite thickness.}
   {No systematic offsets between the observed UV and CO/far-IR emission across the spiral segments are detected. The calculated effective stability parameter has a lowest value of $Q_{\mathrm{eff}} \simeq 1.8$ at galactocentric distances of 12--13 kpc. The least stable wavelengths are rather long, with the lowest values starting from $\simeq 3$~kpc at distances $R > 11$~kpc. }
   {The classical density wave theory is not a realistic explanation for the spiral structure of M\,31. Instead, external causes should be considered, such as interactions with massive gas clouds or dwarf companions of M\,31.}


   \maketitle
%

\section{Introduction}\label{intro}
Density perturbations in galaxy discs often trigger star formation in the form of more or less beautiful spiral structures. An elegant and self-consistent theory of density perturbations as rather weak density waves was proposed by \citet{lin:64}. Using the example of the nearby grand-design spiral galaxy M\,81, \citet{visser:80} concluded that the velocity disturbances seen in the rotation curve are consistent with the density wave theory. As fundamental parameters of the density wave, the pattern speed and corotation radius for M\,81 have been estimated by various authors; the derived results deviate only slightly \citep{feng:14}. For the grand-design galaxy M\,51, \citet{vogel:93} also concluded that the kinematics is in good agreement with the density wave theory (but see a different conclusion by \citet{shetty:07}). \citet{colombo:14} calculated the corotation radii for two- and three-arm models and the pattern speed for the three-arm model of M\,51.  

The pattern speed in a disc of a particular galaxy does not need to be constant, but may vary with galactocentric radius. For example, in the case of M\,51, \citet{meidt:08} derived three different pattern speed values. The density wave theory with a varying pattern speed also allows us to describe more diverse spiral structures. 

However, there are other possible mechanisms to initiate density perturbations in galaxy discs in addition to density waves. Star-forming regions resembling spiral arms may be caused by local instabilities in the disc \citep{goldreich:65, toomre:77, toomre:81, elmegreen:11a, donghia:13, dobbs:14, kumamoto:16} and/or by tidal interactions with satellite galaxies or gas clouds \citep{toomre:72,jaaniste:76,jaaniste:77, oh:08, dobbs:10, pettitt:16, donghia:16}. These mechanisms can be considered as possible sources for density perturbations especially in cases of intermediate or flocculent spirals.

Different scenarios for density perturbations in discs result in different relative distributions of star formation tracers in galaxies. The angular velocity of density waves is generally lower than the angular rotation velocity of the disc. Therefore, if quasi-stationary density waves were responsible for spiral arm formation, spatial offsets should occur between different evolutionary phases of star formation, corresponding to different star formation tracers. Strongest deviations are expected when comparing the distribution of molecular gas or dust densities to the emission from very young stars \citep{roberts:69, tamburro:08, egusa:09, louie:13, hou:15}. From the measurements of such offsets as a function of galactocentric radius, the pattern speed of the density wave can be calculated \citep{egusa:04,louie:13}. This effect causes the pitch angle of spirals to vary with wavelength, which is indeed seen in several galaxies \citep{pour-imani:16}.

On the other hand, if spiral arms were formed due to density inhomogeneities, offsets would not occur, or they would be small and would not have a regular distribution \citep{grand:12a, grand:12b, donghia:13, baba:15, choi:15}. Thus, if no offsets are seen between the distributions of molecular gas and young stars, we may conclude that star formation in the compressed gas has occurred rather fast \citep[see][]{egusa:09}, or that even the classical density wave theory cannot explain spiral structure formation \citep{foyle:11}. 

In the present paper we study the relative distributions of different star formation tracers and derive disc stability conditions in the case of axisymmetric perturbations in the nearby and well-observed Andromeda galaxy (M\,31), which is of de Vaucouleurs type SA(s)b. The spiral structure and star formation rate in M\,31 have been addressed by many authors (see references throughout the paper). For example, the spiral structure has been approximated by two logarithmic spirals predicted by the density wave theory \citep[e.g.][]{braun:91, gordon:06, hu:13}. These spirals are not continuous, but consist of several segments \citep[see also][]{efremov:09}. In addition, a circular ring has been added to the spirals and a more complicated spiral structure has been considered \citep{kirk:15}. In general, as a large and nearby galaxy, M\,31 is a very good target for studying galaxy structure. Moreover, from the point of view of density wave analysis, its non-classical spiral structure with an off-centre ring system makes it different from the much-studied grand-design galaxies referred to above, and in this way, we may touch the possible limits of the density wave theory.

We have assumed the following general parameters for M\,31: a
distance of 785~kpc \citep{mcconnachie:05}, a major axis position angle of $38\degr$ \citep{corbelli:10}, and an inclination angle of $77.5\degr$ \citep{devaucouleurs:91, corbelli:10}. It is known that in the outer parts of the galaxy, the position angle and inclination are not conserved, but vary significantly \citep{chemin:09, corbelli:10}. However, in the present study we consider regions where these parameters can be considered constant.

This paper consists of two main parts and a final section with the results. In the first part, Sect.~\ref{spirals}, we try to detect possible offsets between star formation tracers in M\,31 spirals. In Sect.~\ref{observ} we describe the observational data used for this aim and the method to derive spiral profiles. In Sect.~\ref{correl} the derived correlations are presented. In the second part of the paper, Sect.~\ref{stab}, we study the stability of the disc of M\,31: Sect.~\ref{stab-mod} describes the applied method, Sect.~\ref{stab-param} describes the structural model and the observational data used as the input for the calculations, and Sect.~\ref{stab-res} gives the calculation results. The results are further discussed and conclusions are presented in Sect.~\ref{results}. In the Appendix we present all the figures that we refer to in Sects. \ref{observ} and \ref{correl}.
\section{Distribution of star formation medium and young stars in the M\,31 spirals}\label{spirals}
Star formation processes in galaxies can be studied by analysing the distribution of several star formation tracers: the stellar emission in several optical bands, emission from ionized gas, emission of neutral atomic and molecular gas, and the thermal radiation in the far-IR (FIR) from interstellar dust. In a first approximation the distribution of some tracers is related by the classical Schmidt or Schmidt-Kennicutt relations \citep{schmidt:59, kennicutt:89, tenjes:91, kennicutt:98, braun:09, gonzalez:13, roychowdhury:15}. Here we study the relative distribution of these star formation tracers perpendicular to the spiral arms.  
\subsection{Observational data}\label{observ}
The general stellar mass distribution is well represented by emission in the near-IR (NIR) wavelengths. We have adopted here the $3.6~\mathrm{\mu m}$ data derived by \citet{barmby:06} for mapping the stellar mass of the galaxy.

The number of hot stars is known to correlate well with the far-UV
(FUV) flux, thus \textit{GALEX} photometry is a good characteriser of the distribution of young stars \citep{bianchi:12}. Chemical evolution models of stellar populations indicate that the \textit{GALEX} NUV 2271~\AA\ band corresponds to stellar populations with a mean age of 10~Myr \citep{kennicutt:12}. The FUV 1528~\AA\ band corresponds to the same mean age, but with a smaller age spread. In the present study we have used the near-ultraviolet (NUV) mappings by \citet{thilker:05}.

Before measuring the NUV flux in the regions of interest, we considered the effects of dust extinction. Although high-resolution panchromatic observational data are available for sophisticated starlight-dust interplay modelling \citep[e.g.][]{viaene:17}, the observed UV/FIR ratio is still not recovered perfectly by the models, leaving the actual level on extinction uncertain. In general, since the FIR flux scales with the absorbed optical and UV photons, it is possible to use FIR imaging as a map of dust extinction. The calibration of a FIR map to actual extinction values can be achieved for instance by symmetry considerations \citep{tempel:10,tempel:11}. In the present work we wished to probe the realistic upper and lower limits of extinction to avoid potential systematic effects arising from incorrectly recovered UV flux. To this end, we scaled the \textit{Herschel }$250~\mathrm{\mu m}$ map \citep{fritz:12} as the map of extinction for two extremal cases: to fill the dark patches seen in the FUV maps of the star-forming regions, and to fill the gaps between the spiral arms. The former provides the lower limit, the latter the upper limit of extinction. With very high probability, the true extinction level remains somewhere in between.

At the general metallicity level of M\,31, stars are mainly formed from molecular clouds, although all the details are not clear yet \citep{elmegreen:07,glover:12}. Thus we used the molecular gas distribution determined from CO emission observations by \citet{nieten:06} as the main characteriser of the star formation medium. As complementary data, \ion{H}{i} emission observations by \citet{corbelli:10} and FIR emission measured by \citet{fritz:12} were also used.

All these emission maps of M\,31 were resampled to the pixel scale of 4 arcsec (corresponding to 15~pc). The images were then de-projected to the face-on orientation according to the general parameters listed at the end of Sect.~\ref{intro}.

We decided to compare the relative distribution of different star formation tracers within certain spiral segments, thereby avoiding the necessity for adopting an overall spiral structure model (two-, three- or more-armed structure with corresponding pitch angles) and also decreasing the uncertainties due to possible small variations of the inclination angle of the galaxy with radius.  

Spiral arms or spiral segments can be visually identified from galaxy images or by using some automated method. For example, \citet{egusa:04,egusa:09,egusa:17} determined the peaks in the distribution of star formation tracers by eye. Automated methods include the fitting of Gaussian profiles \citep{dobbs:10}, the cross-correlation method \citep{foyle:11}, etc. Automated methods have their advantages and disadvantages. Offsets derived by using the cross-correlation method can be rather different even when using the same datasets (e.g. results derived by \citealt{foyle:11} and \citealt{tamburro:08}), since they depend on the fitting details, while offsets derived by the Gaussian profile fitting assume symmetric profiles. \citet{louie:13} and \citet{egusa:17}  compared spiral segment selection by eye (distribution peaks), and automated methods for M\,51 and found that both methods give consistent arm locations and that the results are not biased. While testing a new automated method for finding spiral arm segments, \citet{davis:14} verified the results with the spirals found by eye. Thus both methods, the visual selection of spiral segments and detecting them by some automated method, can be justified. Spirals in M\,31 have rather varying widths and symmetry properties. For this reason, we decided to identify spirals segments by eye. 

\begin{figure}
\includegraphics[width=88mm]{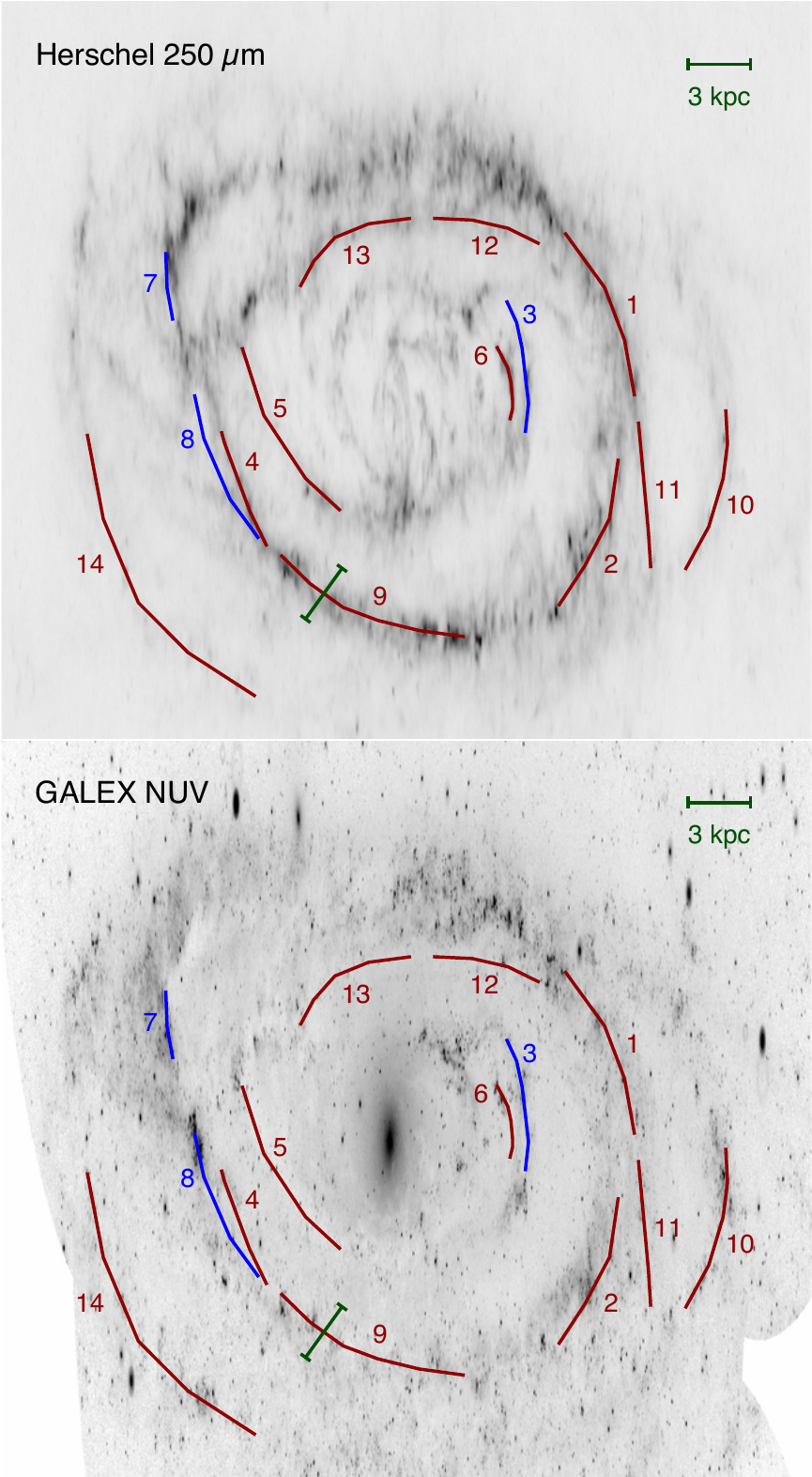}
\caption{De-projected \textit{Herschel} FIR emission map (upper panel) and \textit{GALEX} NUV emission map (lower panel) of M\,31, overlaid with the studied segments. Segments~3, 7 and 8 (blue curves) were defined on the basis of the NUV emission map, the other segments (red curves) from the FIR map. Emission along the segments was integrated, yielding the mean cross section (profiles) of each segment. The width of the cross sections (marked with green line for segment 9) is 3~kpc for all segments.}
\label{f1}
\end{figure}

Fourteen spiral segments were visually selected from the face-on image maps (Fig.~\ref{f1}). Segments 1, 2, 4--6, and 9--14 were selected on the basis of the \textit{Herschel} 250~$\mu$m map, segments 3, 7, and 8 were selected on the basis of the \textit{GALEX} NUV map. Each segment spine was defined as a sequence of connected straight elements. The width of each spiral segment was taken
as 3 kpc, centred on the spine. Thus the segments are wide enough to include the peaks as well as the edges of the spiral arm cross-section profiles.

Emission distributions (profiles) along the cross sections of the spiral segments in each data map were calculated by averaging the intensities along the segments. To further reduce noise that
is due to the clumpy nature of the spiral arms, we masked out small regions that showed up as bright clumps in the \textit{GALEX} NUV image. The same mask was applied to all image maps. For each segment and image map we carried out the following procedure. First, we calculated distances of individual pixels to the spine of the segment. We then grouped pixels into 201 bins of 15~pc width, with bin midpoints located at distances of $i \times 15$~pc ($i = -100\ldots 100$) from the centre line. Negative distances correspond to the inner side of the segment. In each bin, the mean pixel value was calculated and divided by the mean value of all pixels in all bins, yielding a profile for the segment.

\subsection{Correlations between star formation tracers}\label{correl}
One aim of the present study is to compare the spatial distributions of various tracers of different star formation epochs to each other. This has been done using the spiral arm segments and data described in the previous subsection.

\begin{figure}[]
\includegraphics[width=88mm]{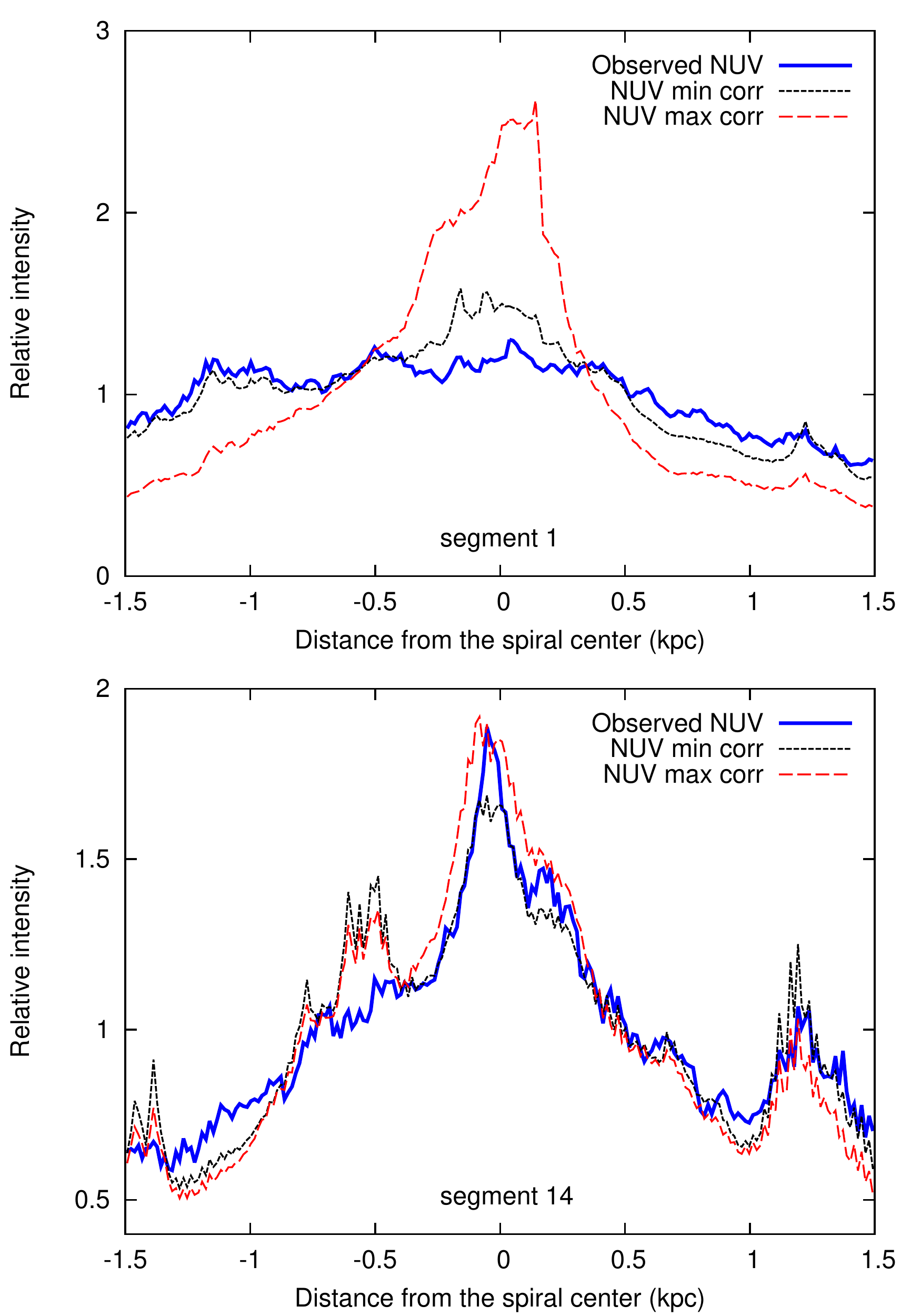}
\caption{NUV emission distribution along the cross sections of spiral arm segments~1 (upper panel) and 14 (lower panel). The blue line denotes emission derived from the original NUV map. Red and black lines denote the upper and lower estimates for the extinction-corrected distribution, respectively; the true extinction-corrected NUV emission distribution lies somewhere in between. The vertical scale is relative.}
\label{f2}
\end{figure}

As mentioned above, we corrected the \textit{GALEX} NUV emission data from dust extinction before further analysis. For most of the segments, the corrections were significant, as illustrated in the upper panel of Fig.~\ref{f2}. Here the original NUV emission distribution along the mean cross section of segment 1 is given with the blue line and the upper and lower estimates for the extinction-corrected distribution are given with the red and green lines, respectively. However, there were also some cases where the extinction corrections were small (e.g. segment 14 shown in the lower panel of Fig.~\ref{f2}). The corresponding NUV profiles for all the segments are given in Fig.~\ref{a1}.

\begin{figure*}
		\center 
        \includegraphics[width=175mm]{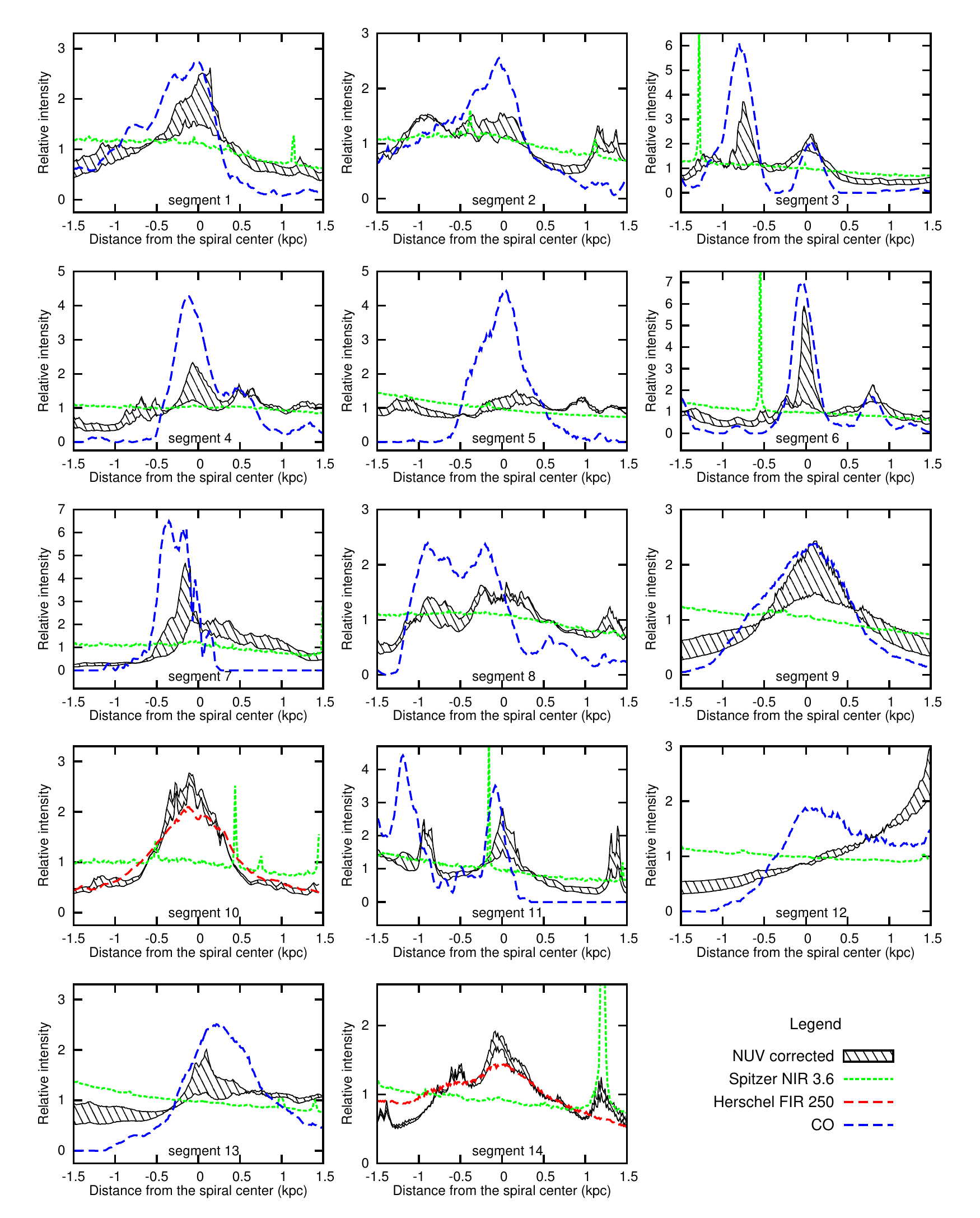}
        \caption{Cross sections of CO emission (or FIR $250~\mathrm{\mu m}$ emission), extinction-corrected \textit{GALEX} NUV emission (young stars with a mean age of 10~Myr; the hatched region corresponds to lower and upper limits of extinction) and \textit{Spitzer }$3.6~\mathrm{\mu m}$ NIR emission (characterising the overall stellar mass distribution) for the spiral segments. The horizontal axis shows the distance from the spine of the given spiral segment. Negative distances correspond to the inner side of a segment (closer to the galactic centre). The vertical axis shows the relative intensity of the corresponding star formation tracer. Narrow peaks in NIR emission distributions are due to local stars.}
\label{f3}
\end{figure*}

To study the dynamics of star formation within the spiral arms, the most telling comparison is between the distribution of young stars and the star formation medium. Although we also measured the respective distributions for \ion{H}{i}, FIR $250~\mathrm{\mu m}$ emission, SDSS g-filter, and \textit{GALEX} FUV emissions, in Fig.~\ref{f3} we present comparison plots only for the extinction-corrected NUV emission (dominated by stars with a mean age of 10~Myr), CO emission (cold gas), and \textit{Spitzer} $3.6~\mathrm{\mu m}$  NIR (characterising the overall stellar mass) profiles. Distribution of FIR $250~\mathrm{\mu m}$, \ion{H}{i}, and CO emissions are given in the Appendix~\ref{app:1}. In these figures, the horizontal axis gives the distance from the spine of the given segment, that is, from the centre of the cross section of the assumed spiral arm. Negative distances correspond to the inner sides of the segments (closer to the galactic centre). On the vertical axis, the relative intensity of the corresponding star formation tracer is shown. The distributions are calibrated with a multiplication factor, keeping the area under the curves equal. 

Our main aim in this section is to find possible offset between the distributions of young stars and the star formation medium in spiral arm segments. According to \citet{braun:91}, the corotation radius in M\,31 is 16.3~kpc, and all the segments discussed here are located inside this radius. When taking the M\,31 pattern speed value as $\Omega_p = 15\, \mathrm{km\, s^{-1} kpc^{-1}}$ \citep{braun:91}, the spiral arms pitch angle value as $9\degr$ \citep{gordon:06,kirk:15}, an average giant molecular cloud lifetime as 25~Myr \citep{blitz:07,fukui:10}, and gas rotation velocities measured by \citet{chemin:09}, the offsets predicted by the density wave theory at galactocentric distances $R = 5-13\,\mathrm{kpc}$ are 0.25--0.5~kpc (see for example \citet{tamburro:08} Eq.~(8) and for the geometrical transformation, their Fig.~1). These predicted offsets can be compared to offsets derived from real emission distribution peaks. Errors of peak positions in the peak-finding method were estimated by \citet{egusa:17} and found to be $\sigma = 0.075\cdot \mathrm{FWHM}$ (the full width at half maximim, $\mathrm{FWHM,}$ corresponds to the emission distribution of a particular profile). By adding the corresponding error estimates for the CO profiles and the NUV profiles presented in Fig.~\ref{f3}, we estimated that typical observed offset errors are 0.03--0.06 kpc. Thus the offsets predicted by the density wave theory should be detectable when comparing emission distributions in M\,31 spiral segments.

In the following, we briefly comment on the emission distribution  in each segment. 

Figure~\ref{f3} shows that emission maxima of segment~1 are quite well defined both in NUV and CO, and the distributions seem to be offset by 0.2~kpc from each other. On the other hand, Fig.~\ref{f1} shows that segment~1 is very likely a part of a major ring-like structure, but it is also possible to interpret it as a spiral arm together with segment~11. The latter is much less pronounced, but has a similar offset by about 0.1~kpc in Fig.~\ref{f3}. However, the predicted offset for segment 1 is 0.35~kpc, for segment~11 it is 0.25~kpc.

Segment~2 can also be interpreted as a possible continuation of segment~1 according to Fig.~\ref{f1}. This segment was defined on the basis of the FIR map, but is even more prominent in the CO map. Comparing the CO profile with the uncorrected NUV profile, an offset by about 0.8~kpc can be seen. This value is too high to be attributed to a usual density wave offset. When taking into account attenuation due to dust (the corrected NUV profile), however, a broad secondary maximum emerges in the NUV distribution with no offset with respect to CO. The predicted offset for segment~2 is 0.35~kpc.

Segments~3 and 6 are quite close to each other and nearly parallel. Segment~3 was selected on the basis of NUV emission distribution, segment~6 on the basis of FIR emission. Although both of them are seen in the corresponding panels in Fig.~\ref{f3}, the distributions are slightly different because averaging has been conducted along different paths. It is not clear whether they form a single spiral. No offsets are seen for either of these segments. The predicted offsets for both segments are 0.5~kpc.

Segment~4 is rather well seen in the FIR and CO maps, but only marginally detected in the NUV map (Fig.~\ref{f1}). No significant offset is detected. The same is valid for segment~5. The predicted offsets for both segments are 0.5~kpc.

Segment 7 is very short, with a length of about 3~kpc. An offset of about 0.3~kpc is seen in Fig.~\ref{f3} (assuming that segment lies in the galaxy plane). This coincides with the predicted offset. The morphology, especially the 3D configuration of this part of the galaxy, is quite unclear. 

Segment 8 is rather uncertain. Figure~\ref{f1} shows that at one end of the segment there is clear and strong NUV emission but no CO emission, while in the part where CO emission is detected no NUV is seen. Therefore, integration of emission along the segment does not allow us to draw any conclusions. The predicted offset for segment~8 is 0.45~kpc.

Segment 9 is well defined both in CO and NUV maps and in the cross-section plot. However, no offset can be detected. The predicted offset for segment~9 is 0.4~kpc.
 
Segments 10 and 14 are the farthest from the galaxy centre and seen in both FIR and NUV. Unfortunately, CO observations are very weak for segment~14 and do not extend to the distance of segment~10. Thus instead of CO, FIR profiles are plotted in the corresponding panels of Fig.~\ref{f3}. Extinction corrections have been small for these segments, allowing us to compare the NUV emission to the FIR emission. Both segments are rather close to the corotation radius estimated by \citet{braun:91}, thus there should be no offsets, and indeed no offsets are visible. However, no offset should be visible without the classical density wave theory as well. 

Segment 12 is seen in CO and FIR but not in NUV, and an offset cannot be detected. According to CO and FIR, segment~13 may be a continuation of 12. In the extinction-corrected NUV profile a marginal NUV enhancement can be seen with no significant offset, but no clear conclusions can be formulated. The predicted offsets for both segments are about 0.45~kpc.

For all the segments, \textit{Spitzer} 3.6~$\mu$m emission profiles were also calculated in order to find possible mass surface density enhancements within the traced structures. As Fig.~\ref{f3} shows, no significant enhancements are seen. 
\section{Disc stability analysis}\label{stab}
Since we did not find offsets between the distributions of the star-forming medium and young stars predicted by the classical density wave theory, we decided to study the dynamical state of the disc components, in particular their stability to perturbations. For this we needed a mass distribution model of M\,31, which we took from \citet{tamm:12}. In this model it was assumed that the galaxy is in a stationary state and can be approximated as a superposition of several axisymmetric components that represent actual stellar populations. The components are approximated with ellipsoidal mass distributions with different ellipticities. 

The disc part of M\,31 consists of several gaseous and stellar components with different composition and properties. All these flattened components are also affected by the external gravitational potential of the spheroidal part of M\,31 consisting of a bulge, a stellar halo, and a dark matter halo. Our model has to take all these aspects into account. 
\subsection{Stability model}\label{stab-mod}
A proper study of galaxy stability has to take into account all these structural properties. The first models studying the resilience of a system to axisymmetric perturbations, considering both stellar and gas disc components of the galaxy, were developed by \citet{jog:84}, \citet{romeo:92}, and \citet{elmegreen:95}. These models also take into account the finite thickness of both discs; the perturbation theory for a stellar disc with a finite thickness was thoroughly studied earlier by \citet{vandervoort:70}. Perturbation models for a galaxy consisting of several stellar disc components and a gas disc were thereafter studied by \citet{rafikov:01}, who derived a quite easily applicable condition for the stability of such a system, which is valid for highly flattened discs. Perturbation models generalised for several gas disc components were developed by \citet{romeo:13}. 

According to \citet{vandervoort:70} and \citet{romeo:92}, accounting for the finite thickness of a component reduces to a factor $(1 + k h_{\mathrm{eff}})^{-1}$, where $k$ is the wavenumber of the perturbation and $h_{\mathrm{eff}}$ is the effective scale height of a component. Such a thickness correction was demonstrated to be quite accurate \citep{elmegreen:11a}.

Taking the above mentioned correction for finite thickness into account, the stability condition (or the stability curve $S\!C (k,R)$) as a function of wavenumber $k$ and galactocentric radius $R$ can be written in the form \citep[see][]{romeo:92,rafikov:01,romeo:13}
\[
S\!C(k,R) = \sum_{i=1}^n   \frac{2\pi G}{k \sigma_i^2 (R)} \left[ 1-e^{q_i^2 (R)} I_0 (q_i^2 (R)) \right] S_i (R) \cdot \frac{1}{(1+k \, h_{i,\mathrm{eff}} ) } +
\]      
\begin{equation}
 +  \sum_{j=1}^m   \frac{2\pi Gk}{\kappa^2 (R) + k^2 c_j^2 (R)} S _j (R) \cdot \frac{1}{(1+k \, h_{j,\mathrm{eff}} ) }    < 1.  \label{eq:stab}
\end{equation}
The first sum (summation index $i$) in this condition accounts for stellar disc components and the second sum (index $j$) for gaseous disc components. The quantity $q$ is defined as in \citet{romeo:92} and \citet{rafikov:01} 
\begin{equation}
q_i^2(R) = k^2 \sigma_i^2 (R) /\kappa^2 (R)
\label{eq:q}
\end{equation}
and the effective scale height as in \citet{romeo:92} 
\begin{equation}
h_{\mathrm{eff}} = S (R) /(2\rho(R,0)).
\label{eq:h}
\end{equation}
In Eqs. (\ref{eq:stab}) -- (\ref{eq:h}), $G$ is the gravitational constant, $\sigma$ is the velocity dispersion of a given stellar disc component in the $R$ direction, $c$ is the speed of sound in a given gaseous component, $I_0$ is the modified zero-order Bessel function, $\kappa$ is the epicyclic frequency of the galaxy, and $S$ and $\rho$ are the mass surface density and volume density of the component, respectively. 

The contribution of the external gravitational potential by the bulge, the stellar halo, and the dark matter halo to stability condition is included in the epicyclic frequency $\kappa^2 (R) = \frac{\partial}{\partial R} K_R + \frac{3}{R}K_R $, where $K_R$ is the total gravitational potential derivative in $R$ direction \citep[see also][]{jog:14}.

From calculated stability condition curves $S\!C(k,R)$ it is easy to derive effective stability parameters $Q_{\mathrm{eff}}$ defined as \citep[see e.g.][]{romeo:13}
\begin{equation}
        Q_{\mathrm{eff}}^{-1} (R) =\max_k ~\{ S\!C(k,R) \} 
        \label{eq:qeff}
\end{equation}
and the corresponding least stable perturbation wavenumber $k_{\ast}$ or wavelength $\lambda_{\ast} = 2\pi /k_{\ast} $ values. \citet{romeo:11} and \citet{romeo:13} derived easier to use expressions for the stability parameter $Q_{\mathrm{eff}}$, but our intention is to study also the contribution of individual disc components to the overall stability and to estimate the stability condition as a function of the perturbation wavelength, thus we use condition (\ref{eq:stab}).

\begin{table*}
        \caption[]{Mass distribution parameters for the disc components of M\,31 (see Sect.~\ref{stab-param}).}      
        \label{tab:mud}
    \centering
        \begin{tabular}{lcccccc}
        \hline\hline    
Component &     $a_c$ & $\epsilon$ & $N$ & $d_N$ &      $\rho_c$ & $M$  \\
                  & [kpc] &     &     &       & $[\mathrm{M_{\sun}\,pc^{-3}}]$   & 
                  $[10^{9}\,\mathrm{M_{\sun}}]$ \\
        \hline
Main stellar disc  & 10.67 & 0.17 & 1.2 & 3.273 & $1.307\cdot 10^{-2}$ & 56 \\
Young stellar disc & 11.83 & 0.01 & 0.2 & 0.316 & $1.179\cdot 10^{-2}$ & 1 \\ 
Cold gas disc      & 11.83 & 0.01 & 0.2 & 0.316 & $2.946\cdot 10^{-2}$ & 2.5\\
Warm gas disc      & 11.83 & 0.03 & 0.2 & 0.316 & $1.375\cdot 10^{-2}$ & 3.5\\
\hline
\end{tabular}
\end{table*}
\subsection{M\,31 model parameters}\label{stab-param}
To calculate the stability parameter of M\,31, we took the mass distribution model published in \citet{tamm:12}. In this stationary axisymmetric model a galaxy is assumed to consist of several components. Density distributions of the components are given by Einasto's formula
\begin{equation}
\rho (a) = \rho_c\exp\left\{-d_{N}
\left[\left( a / a_\mathrm{c} \right)^{1/N}-1\right]\right\} .
\label{eq:einasto}
\end{equation}
Here the distance from the centre $a=\sqrt{R^2+z^2/\epsilon^2}$, where $R$ and $z$ are two cylindrical coordinates; $\epsilon$ is the axial ratio, $N$ is a structure parameter, and $d_N$ is a function of $N$, such that $\rho_c$ becomes the density at radius $a_{\mathrm{c}}$, which defines a volume containing half of the total mass of the component \citep[for details, see][]{tamm:12}. 

The gravitational potential derivative of a component, necessary for calculating epicyclic frequencies, can in this case be expressed in a convenient form \citep[see e.g.][]{tenjes:01}
\begin{equation}
        K_R (R,z) \equiv \frac{\partial\Phi}{\partial R} = R \frac{4\pi \epsilon G}{e^3}\int_0^{\arcsin e} \rho (a) \sin x\, \mathrm{d}x ,
        \label{eq:grpot}
\end{equation}
where $e=\sqrt{1-\epsilon^2}$ is eccentricity and $a^2 = \frac{\sin^2 x}{e^2} \left( R^2 + \frac{z^2}{\cos^2 x} \right)$. The formula is general, but the epicyclic frequencies were calculated within the galactic plane only. 

While the mass distribution parameters of the spheroidal components and the main stellar disc can be used as presented in \citet{tamm:12}, the flat young disc needs to be split further into main subsystems. At this stage we do not yet know the relative contributions of different flat components to the overall galaxy stability, thus we have to analyse each one of them. 

First we separated gas and stars by taking the mass of young stars to be $1 \cdot 10^{9}\,\mathrm{M_{\sun}}$ and the gas mass to be $6 \cdot 10^{9}\,\mathrm{M_{\sun}}$ \citep{tamm:12}. Thus we now had two stellar discs, one containing old and intermediate-age stars, and the other containing young stars. The parameters of these flat stellar components are given in Table \ref{tab:mud}. 

Several subcomponents can also be distinguished in the gas disc. Although four to five different phases can be distinguished in the interstellar medium \citep{mckee:77,chemin:09}, here we have considered the neutral gas (including both the atomic and the molecular component) to consist of two main phases, the cold neutral medium (CNM) and the warm neutral medium (WNM) \citep[see][]{brinks:84,wolfire:03,kalberla:09}. For M\,31 \citet{dickey:93} estimated that the mass of WNM is about 60 per cent and the mass of CNM is 40 per cent of the total neutral gas mass. These proportions have been found to also be valid in the Milky Way: according to \citet{heiles:03}, the mass of WNM is 60 per cent of the total neutral gas mass. Thus we took the masses $M_{\mathrm{WNM}} = 3.5\cdot 10^{9}\,\mathrm{M_{\sun}}$ and $M_{\mathrm{CNM}} = 2.5\cdot 10^{9}\,\mathrm{M_{\sun}}$.

Unfortunately, direct measurements of the thickness of these two gas components in M\,31 do not exist. \citet{brinks:84} estimated from their gas velocity dispersion measurements that CNM has $\epsilon\simeq 0.01$. In addition, taking into account the broad similarity between M\,31 and the Milky Way, we have applied here the estimate by \citet{kalberla:03} that at the solar distance, the WNM disc of the Milky Way is about three times thicker than the CNM disc. Thus we take $\epsilon_{\mathrm{CNM}} = 0.01$ and $\epsilon_{\mathrm{WNM}} = 0.03$. However, to consider the impact of uncertainties, we also performed calculations for some other flatnesses. All the other structure parameters and effective radii were kept unchanged. The parameters of the two gas discs are given in Table~\ref{tab:mud}.

To complete the set of input parameters for stability calculations, we also needed to know velocity dispersions and the speed of sound in the disc components. Following \citet{dorman:15}, we took velocity dispersions of the main disc stars at distances 7--15~kpc to be $25-40~\mathrm{km\,s^{-1}}$, at 5~kpc we took an extrapolated value $40~\mathrm{km\,s^{-1}}$. \citet{aniyan:16} estimated that velocity dispersion of young disc stars is about a half of this; we applied a constant value $15~\mathrm{km\,s^{-1}}$ for the young stellar component, which is also consistent with an earlier estimate by \citet{tabatabaei:10}. However, since the velocity dispersion of the young disc is an uncertain quantity, we also calculated dispersions of $10~\mathrm{km\,s^{-1}}$ and $22.5~\mathrm{km\,s^{-1}}$.

Sound speed in gas discs can be calculated from gas temperature. According to \citet{dickey:93}, \citet{heiles:01}, and \citet{braun:09}, the average CNM spin temperature can be taken as $T_S \sim 50$~K, giving for the thermal sound speed $c_T = 0.6~\mathrm{km\,s^{-1}}$. However, other factors have also to be taken into account, for
example, turbulences within gaseous media. In this case we have the square sum of all the contributing components in Eq.~(\ref{eq:stab}) instead of a simple thermal sound speed \citep{elmegreen:11a, hoffmann:12}. In the Milky Way, the gas disc Alfv\'en velocity due to magnetic field pressure is $v_A = 1.5~\mathrm{km\,s^{-1}}$ \citep{heiles:05}. Since the average magnetic field induction in M\,31 \citep[see][]{fletcher:04} and the value used by \citet{heiles:05} are similar, we took the $v_A$ value of the Milky Way to be also
representative of M\,31. Velocities corresponding to the turbulence velocities in M\,31 are not directly known, thus we use a typical value of $v_{\mathrm{turb}} = 6~\mathrm{km\,s^{-1}}$ \citep{kennicutt:89,martin:01,romeo:13,lee:16} as the first approximation. Comparing the thermal sound speed, Alfv\'en velocities, and turbulent velocities, it is clear that turbulent velocities dominate, and we take for our calculations $c_{\mathrm{CNM}} = 6~\mathrm{km\,s^{-1}}$. 

The WNM temperature is even more uncertain. For example, \citet{heiles:01} and \citet{braun:09} have suggested that WNM temperature observations are uncertain because of non-thermal broadening. Following the theoretical model by \citet{wolfire:95, wolfire:03}, we have taken WNM temperature to be $T \sim 8000$~K, corresponding to the speed of light $c_s \sim 7.3~\mathrm{km\,s^{-1}}$. 

The velocity parameters of the disc components used for our calculations are given in Table \ref{tab:disp}.
\subsection{Stability parameter calculations}\label{stab-res}
\begin{figure}
\includegraphics[width=88mm]{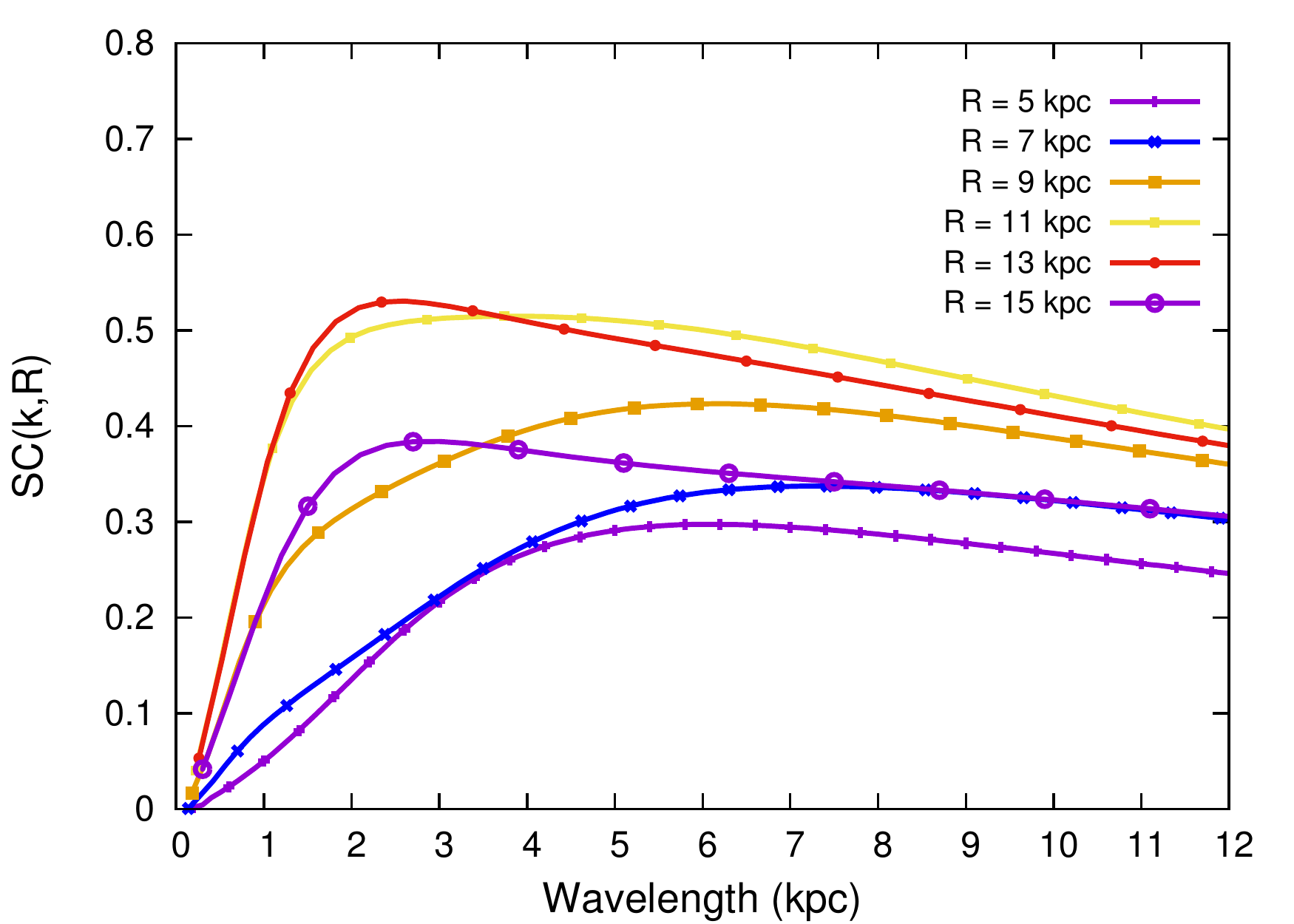}
\caption{Calculated stability curves $SC(k,R)$ as a function of perturbation wavelength at various galactocentric distances $R$.}
\label{f4}
\end{figure}

Using the parameters given in Tables~\ref{tab:mud} and \ref{tab:disp} and the external gravitational potential according to \citet{tamm:12}, we calculated stability curves $S\!C(k,R)$ as a function of perturbation wavelength at galactocentric distances from R = 5 to 17~kpc according to Eqs. (\ref{eq:stab})--(\ref{eq:h}). The calculated curves for R = 5, 7, 9, 11, 13, and 15~kpc are plotted in Fig.~\ref{f4}. 

\begin{figure}
\includegraphics[width=88mm]{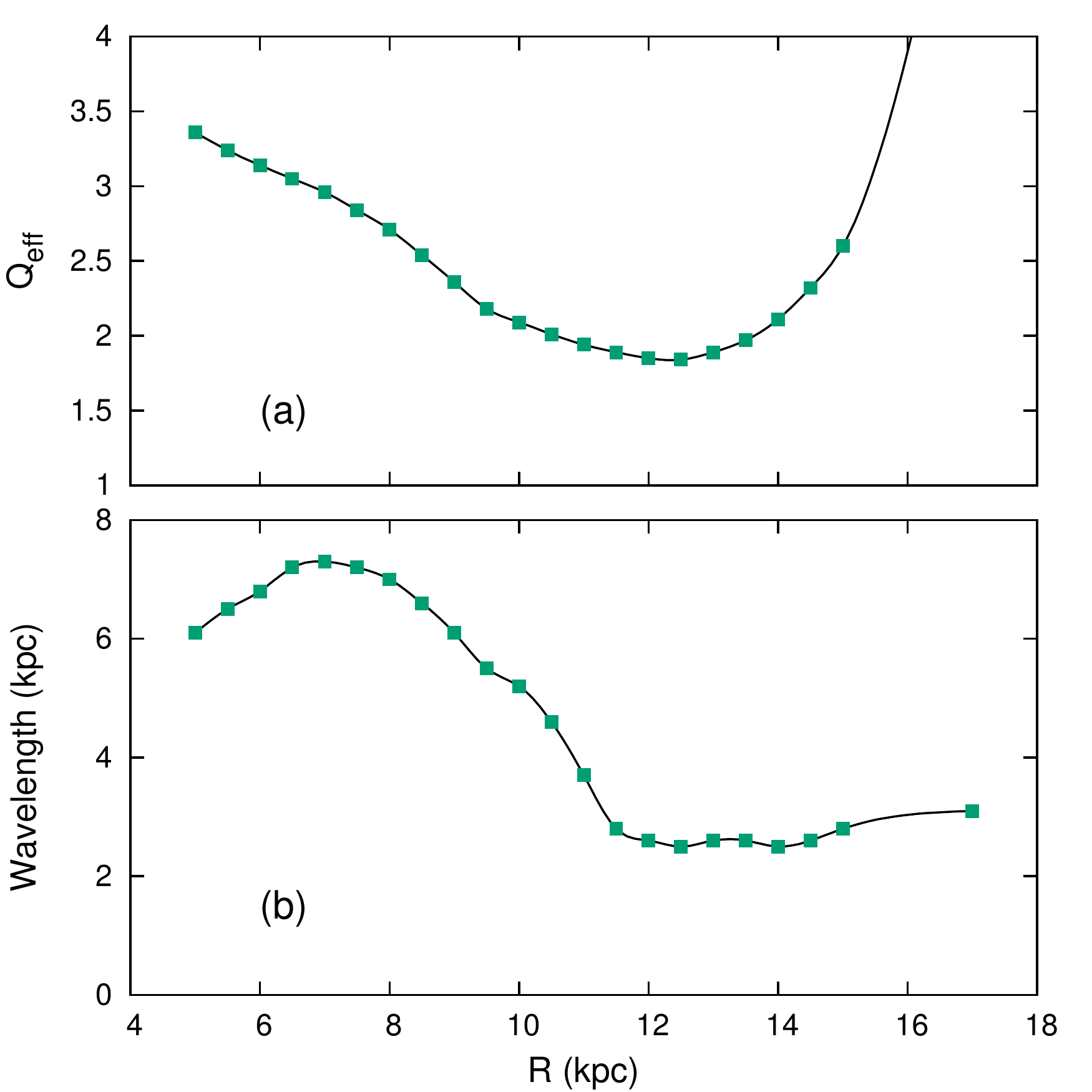}
\caption{Calculated stability parameter $Q_{\mathrm{eff}}$ (upper panel) and the least stable wavelength $\lambda_{\ast}$ (lower panel) as functions of the galactocentric distance $R$.}
\label{f5}
\end{figure}

Effective stability parameters $Q_{\mathrm{eff}}$ derived from the calculated stability curves according to Eq.~(\ref{eq:qeff}) and the corresponding least stable perturbation wavelengths $\lambda_{\ast}$ as functions of $R$ are presented in Fig.~\ref{f5}.

\begin{figure}
        \includegraphics[width=88mm]{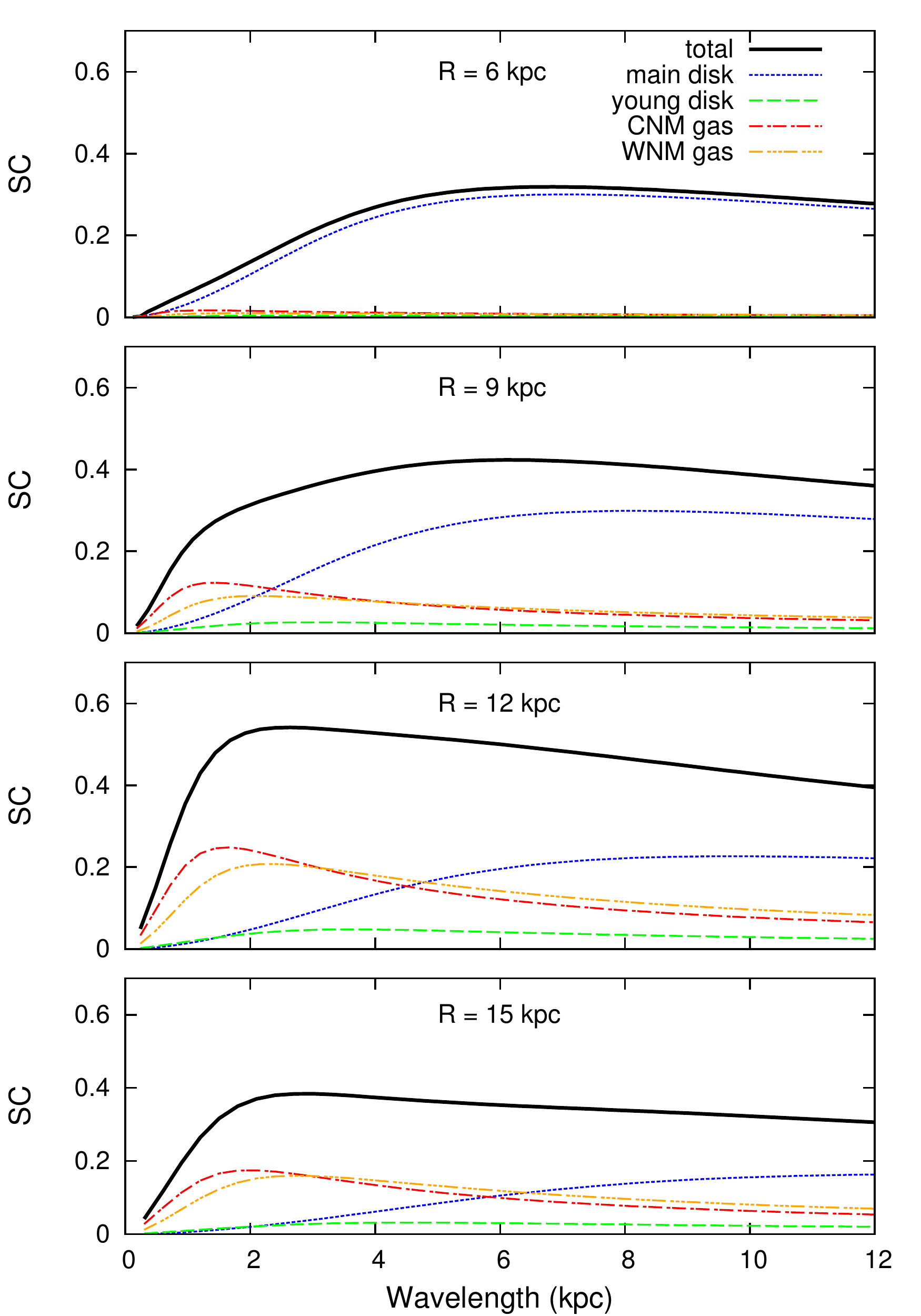}
        \caption{Calculated individual terms in stability curve $S\!C$ for all stellar and gas disc components at galactocentric distances $R = $ 6, 9, 12, and 15~kpc as a function of the perturbation wavelength.}
\label{f6}
\end{figure}

Figure~\ref{f6} shows the contribution of individual disc components to the total stability curves (\ref{eq:stab}). At $R=6$~kpc the contribution of the main stellar disc dominates for wavelengths $\lambda > 0.5$~kpc. At perturbation scales $\lambda < 0.5$~kpc the cold gas disc is also important. At larger radii, both cold and hot gas become important in determining the stability of the galaxy, especially at shorter perturbation wavelengths. At $R=9$~kpc the stellar disc starts to dominate beyond perturbation wavelengths $\lambda > 2$~kpc. At distances $R = 12$ and 15~kpc the stellar disc and the cold and hot gas discs give a comparable contribution to the stability criterion (\ref{eq:stab}). The young disc has a minor impact on stability curves at all wavelengths and radii.

\begin{table}
        \caption[]{Velocity distribution parameters for the disc components of M\,31.}   
        \label{tab:disp}
        \begin{flushleft}
        \center 
        \begin{tabular}{lll}
        \hline\hline    
Component          & $\sigma$ or $c$       & Comments  \\
                           & $\mathrm{km\,s^{-1}}$ &     \\
        \hline
Main stellar disc  & 25--40                & a function of $R$ \\
Young stellar disc & 15\tablefootmark{a}   & a constant value \\      
Cold gas disc      & 6                     & a constant value \\
Warm gas disc      & 7.3                   & a constant value \\
\hline
\end{tabular}
\tablefoot{ \tablefoottext{a}{Calculations were also made for dispersions of $10~\mathrm{km\,s^{-1}}$ and $22.5~\mathrm{km\,s^{-1}}$.} }
\end{flushleft}
\end{table}

We described above that some parameter values we used as the input data are known only approximately. The results presented in Figs.~\ref{f4} and \ref{f6} were reached by assuming the velocity dispersion of the young stellar disc to be $15~\mathrm{km\,s^{-1}}$. Performing similar calculations for dispersion values $7.5~\mathrm{km\,s^{-1}}$ and $22.5~\mathrm{km\,s^{-1}}$, we found that for smaller scale (0.5--4~kpc) perturbations $Q_{\mathrm{eff}}$ changes by less than 10 per cent. For larger scale perturbations the changes were even smaller. 

Similarly, since the flatness of the CNM gas disc is not well known either, we also varied $\epsilon_{\mathrm{CNM}}$ (note that while varying the CNM disc flatness, the gas surface density has to be kept unchanged). When increasing $\epsilon_{\mathrm{CNM}} $, the contribution of CNM disc to $Q_{\mathrm{eff}}$ decreases. At $R=10$~kpc, for example, while taking $\epsilon_{\mathrm{CNM}}= 0.05$, the contribution of CNM decreases about twofold at $\lambda = 3$~kpc and about threefold at $\lambda = 1$~kpc. As a result, $Q_{\mathrm{eff}}$ also increases. 
\section{Results and discussion}\label{results}
On the basis of de-projected face-on images of M\,31, fourteen spiral arm segments were defined. Measurements of \textit{GALEX} NUV, \textit{Herschel} $250~\mu$m FIR, and CO emission was averaged along these segments and the resulting mean cross-section distributions in different datasets were compared to each other, in order to detect possible spatial offsets. For most of the considered segments, no significant offsets between the different star formation tracers were found.

Disc structure analysis and NIR imaging has revealed that spiral arms are not mere tracers of star formation, but typically also enhancements of stellar mass density \citep{schweizer:76, eskridge:02, elmegreen:11b, kendall:11, kendall:15}. We also calculated \textit{Spitzer} 3.6~$\mu$m emission cross-section profiles of the spiral segments, in order to find possible mass surface density enhancements at the locations of the spiral structures of M\,31. No significant enhancements were found (see Fig.~\ref{f3}, green lines). Thus star formation in spiral arms is not caused by density enhancement due to a simple stacking of stellar orbits. 

Disc component stability curve calculations for axisymmetric perturbations yielded that $Q_{\mathrm{eff}} \ge 1.8$ all over the disc. Similar rather high values of $Q_{\mathrm{eff}}$ were derived by \citet{leroy:08}, \citet{romeo:13}, \citet{romeo:15},
and \citet{westfall:14} for several galaxies. Thus, high $Q_{\mathrm{eff}}$ values can be quite common, and galaxy discs are safely stable against local radial perturbations. The least stable wavelengths  (Fig.~\ref{f5}b) are $\lambda_{\ast} > 3$~kpc, being comparable to the distances between several prominent spiral segments. Figure~\ref{f6} shows that when only the cold gas is considered, $\lambda_{\mathrm{\ast, CNM}} = 1.3 - 2$~kpc, which is always shorter than for the whole disc. 

Rather high $Q_{\mathrm{eff}}$ values may indicate that the disc of M\,31 is too stable to support density waves. As was pointed out by \citet{toomre:69, toomre:77}, in the case of high $Q$ values the distance range in a galaxy disc where density waves do not vanish is rather narrow. However, to derive the criterion for the permitted region, Toomre assumed a radially constant $Q$. On the other hand, our calculation of $Q_{\mathrm{eff}}(R)$ for M31 (see Fig.~\ref{f5}a) shows that this assumption is not well applicable. According to \citet{khoperskov:12}, such a U-shaped distribution of $Q$ may favour density waves at a certain range of radii. 

It is essential to mention that the above derived $Q_{\mathrm{eff}}$ calculations are based on the assumption of axisymmetric linear perturbations. At the end of Sect.~\ref{correl} we concluded that no significant mass surface density enhancements are seen in spiral segments and the assumption of small perturbations is satisfied. The model does not take into account gas dissipation and non-axisymmetric perturbations \citep[see][]{romeo:16}, however. For example, when the dissipation is into account, the numerical values of $Q_{\mathrm{eff}}$ might change in a complicated way depending on the perturbation wavelength, thickness of discs, dissipation efficiency, etc. \citep{elmegreen:11a}, and the critical $Q$ value for the instability can even be $Q_{c} = 2-3$. In this case, Fig.~\ref{f5}a shows that over most of the disc, M\,31 disc is close to the stability-instability limit, but without the expected offsets between different star formation tracers.  

For the majority of galaxies, star formation is localised in spiral arms. However, this is not always the case. Star formation may spread throughout the galactic disc and even far outside the main disc, where the surface density of gas is rather low. For example, in M\,33 \citet{grossi:11} discovered star formation even at a distance of about 10 disc scale lengths from the galaxy centre, in NGC\,2403 \citet{barker:12} found a young star at a distance of 8 disc scale lengths. Unfortunately, stability calculations at very large distances from the centre of M\,31 are extremely uncertain as we do not know stellar velocity dispersions, gas geometry, and effective sound speed in these regions. In addition, disc surface density estimates are very uncertain. For this reason we have limited our calculations to $R \le 17$~kpc.

Expected offsets between different star formation tracers are absent in M\,31 spirals. Thus we can conclude that throughout most of the disc of M\,31, density waves would be suppressed and are weak or absent in observations. This result supports N-body/SPH simulations by \citet{grand:12a,grand:12b} and \citet{donghia:13}, indicating that spiral arms may be transient features with variable pattern speeds and without significant offset between different star-forming tracers. According to \citet{sellwood:11}, the transient nature of spirals is also suggested by the disc kinematics in the solar neighbourhood. Very weak or absent density waves in M\,31 were also concluded by \citet{fletcher:04} from magnetic field orientations derived from radio polarization observations. Star formation regions resembling spiral arms may be caused by local instabilities in discs or tidal interactions with satellite galaxies or gas clouds (see references in Sect.~\ref{intro}). Indeed, an extended filament of \ion{H}{i} clouds in the vicinity of M\,31 has been observed \citep{braun:04,lockman:12,wolfe:16}, and data and simulations also suggest that interactions with M\,32, M\,33, and NGC\,205 may have taken place \citep{McConnachie:09,Putman:09,Choi:02,Dierickx:14}, supporting the idea of an external origin of the spiral structure of the Andromeda galaxy.

This study concerns only one particular galaxy with a rather specific spiral structure. We caution against using it as the basis for far-reaching implications about the overall spiral structure formation mechanisms in general.
\begin{acknowledgements}
      We are grateful to the referee, Alessandro Romeo, whose detailed, informative, and constructive comments helped us to greatly to improve the paper. We acknowledge the support by the Estonian Research Council grants IUT26-2, IUT40-2, IUT2-27, and by the European Regional Development Fund (TK133). This research has made use of NASA's Astrophysical Data System Bibliographic Services. IRAF is distributed by the National Optical Astronomy Observatory, which is operated by the Association of Universities for Research in Astronomy (AURA) under a cooperative agreement with the National Science Foundation.
\end{acknowledgements}
\bibliographystyle{aa} 
\begin{appendix}
		\onecolumn 
\section{Emission distributions for the spiral segments~1--14.} \label{app:1}
\begin{figure*}[h]
		\center 
        \includegraphics[width=157mm]{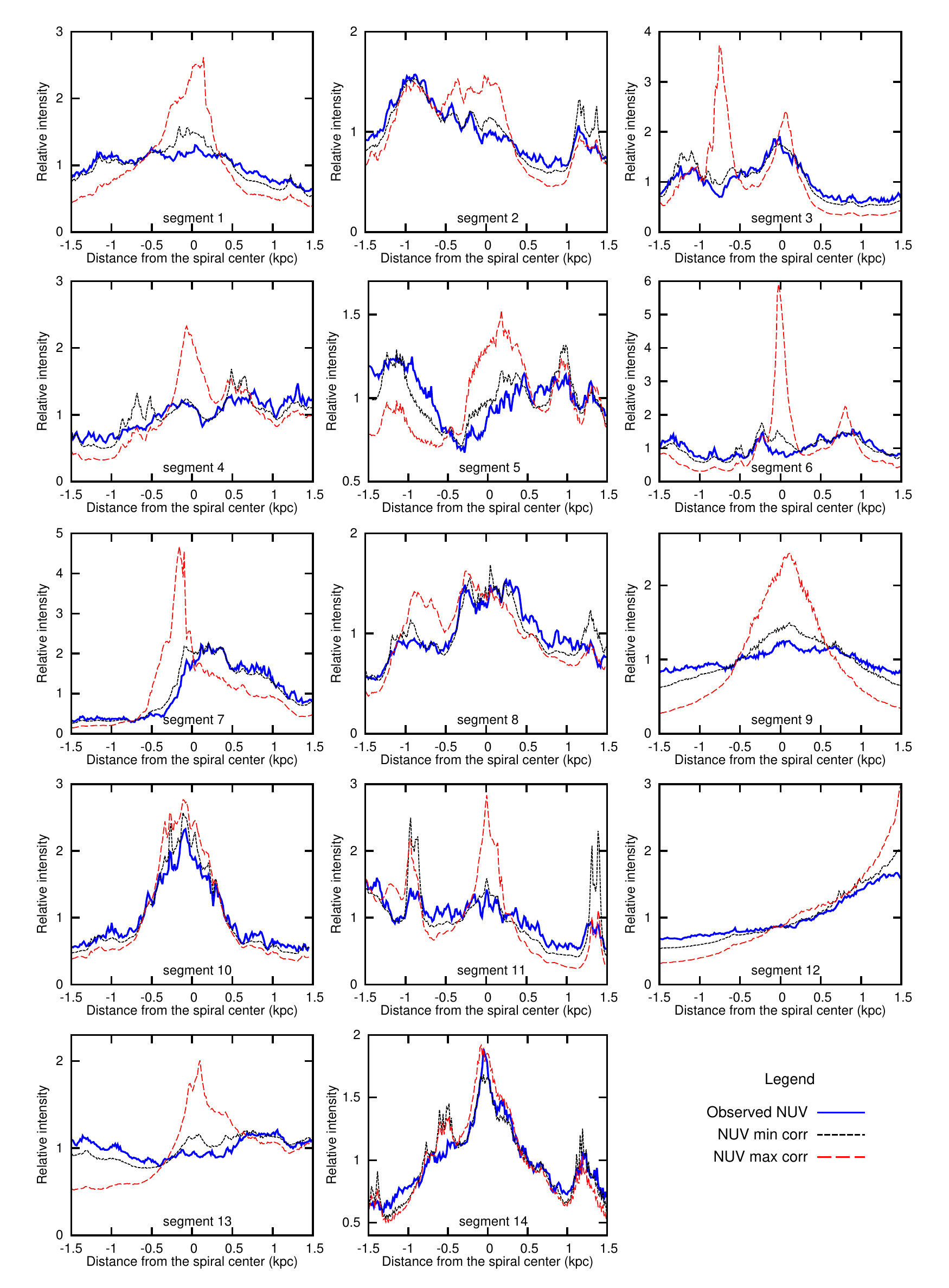}
        \caption{Emission distributions for segments~1--14 derived from the original NUV map (blue line) and extinction-corrected maps. The lowest absorption correction is given by the black dashed line, and the highest correction by the red dotted line. The true absorption-corrected NUV emission distribution lies between these lines.}
\label{a1}
\end{figure*}

\begin{figure*}
		\center 
        \includegraphics[width=170mm]{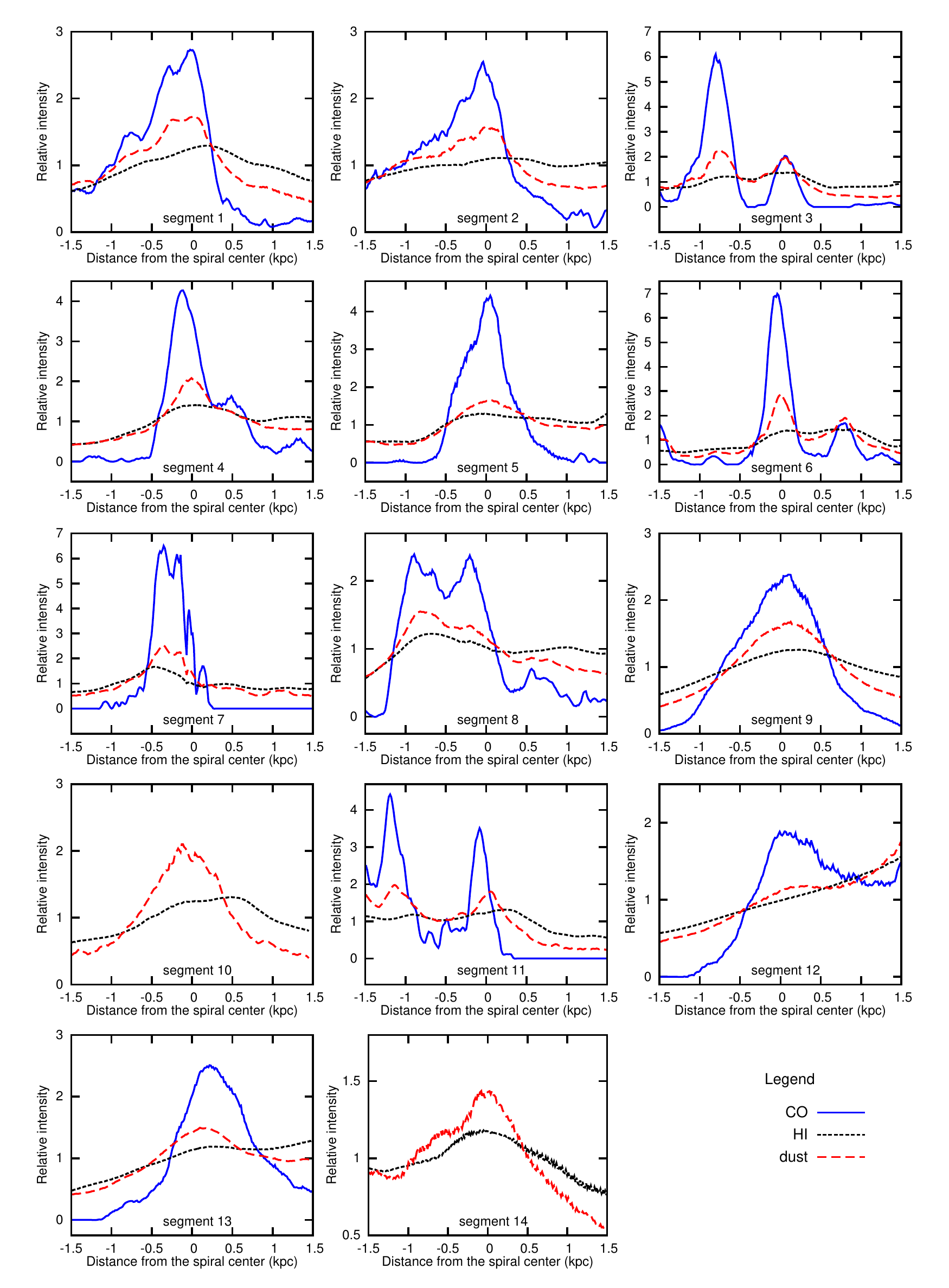}
        \caption{Emission distributions for segments~1--14 derived from the CO emission maps (blue lines), \ion{H}{I} emission maps (black lines), and 250~$\mu$m dust emission maps (red lines). }
\label{a2}
\end{figure*}
\end{appendix}
\end{document}